\documentclass[twocolumn,showpacs,preprintnumbers,amsmath,amssymb]{revtex4}

\usepackage{graphicx}% Include figure files
\usepackage{dcolumn}% Align table columns on decimal point
\usepackage{bm}% bold math
\usepackage{amssymb}
\usepackage{epsfig}    
\usepackage{here}

\newcommand{\beq}{\begin{equation}}
\newcommand{\eeq}{\end{equation}}
\newcommand{\bq}{\begin{equation}}
\newcommand{\eq}{\end{equation}}
\newcommand{\ba}{\begin{array}}
\newcommand{\ea}{\end{array}}
\newcommand{\beqa}{\begin{eqnarray}}
\newcommand{\eeqa}{\end{eqnarray}}

\newcommand{\tx}{\texttt}
\newcommand{\as}{$_{\!\times}$}
\newcommand{\sst}{\mbox{\boldmath $^{\star}$}}

\newcommand{\ZZ}{\mathbb{Z}}
\def\RE{\Re\mathfrak{e}}

\def\TT{{\cal T}}
\def\T{{\cal T}}

\def\Th{\widehat{\cal T}}

\def\End{\end{document}}

\def\to{\rightarrow}

\def\dis{\displaystyle}
\def\f{\frac}

\def\[{\left[}
\def\]{\right]}
\def\({\left(}
\def\){\right)}

\def\pih{\widetilde{\pi}}

\def\l{{\ell}}

\def\a{\alpha}

\def\U1EM{U(1)_{\rm em}}

\def\T{{\cal T}}
\def\leqq{\leqslant}
\def\geqq{\geqslant}

\def\[{\left[}
\def\]{\right]}
\def\dis{\displaystyle}

\def\cut{\Lambda}

\def\d{\delta}
\def\mH{m^{~}_H}
\def\la{\lambda}

\def\lah{\widehat{\lambda}}

\def\gh{\widehat{g}}
\def\gsh{\widehat{g}_s^{~}}
\def\ggsh{\widehat{g}_s^2}
\def\Phih{\widehat{\Phi}}
\def\phih{\widehat{\phi}}
\def\pih{\widehat{\pi}}

\def\HB{\overline{H}}
\def\hh{\widehat{h}}
\def\Wh{\widehat{W}}
\def\Gh{\widehat{G}}
\def\Bh{\widehat{B}}

\def\sq{\sqrt{2}}
\def\vd{\widehat{v}}
\def\nb{\rm{\bf n}}
\def\NKK{N_{\rm KK}}

\def\MG{M_{\rm G}}
\def\a{\alpha}

\def\ainv{\alpha^{-1}}
\def\bt{\widetilde{b}}
\def\k{\kappa}
\def\D{\Delta}

\def\thisday{February, 2003}

\begin{document}

\preprint{\large hep-ph/0302263}

\title{% 
Unitarity of the Higher Dimensional Standard Model}
% Force line breaks with \\

\author{{\sc R. Sekhar Chivukula}\,$^a$,~\,
        {\sc Duane A. Dicus}\,$^b$,~\,
        {\sc Hong-Jian He}\,$^b$,~\,  
        {\sc Satyanarayan Nandi}\,$^c$}
\email{sekhar@bu.edu, dicus@physics.utexas.edu, \linebreak ,       
       hjhe@physics.utexas.edu, shaown@okstate.edu}%    
\affiliation{%
\vspace*{2mm}
$^{a}$\,Department of Physics, Boston University, 
      590 Commonwealth Ave., Boston, MA 02215, USA \\
$^{b}$\,Center for Particle Physics and Department of Physics, 
      University of Texas at Austin, TX 78712, USA\\
$^{c}$\,Department of Physics, Oklahoma State University,
      Stillwater, OK 74078, USA       
}%
\date{\thisday}% It is always \today, today,

\begin{abstract}
\noindent
We study the unitarity of the standard model (SM) in higher
dimensions.  We show that the essential features of SM unitarity
remain after compactification, and place bounds on the highest
Kaluza-Klein (KK) level $\NKK$ and the Higgs mass $\mH$ in the
effective four-dimensional (4d) low-energy theory.  We demonstrate
these general observations by explicitly analyzing the effective 4d KK
theory of a compactified 5d SM on $S^1/\ZZ_2$.  The nontrivial energy
cancellations in the scattering of longitudinal KK gluons or KK weak
bosons, a consequence of the geometric Higgs mechanism, are
verified. In the case of the electroweak gauge bosons, the
longitudinal KK states also include a small mixture from the KK Higgs
excitations.  With the analyses before and after compactification, we
derive the strongest bounds on $\NKK$ from gauge KK scattering.
Applying these bounds to higher-dimensional SUSY GUTs implies that
only a small number of KK states can be used to accelerate gauge
coupling unification. As a consequence, we show that the GUT scale 
in the 5d minimal SUSY GUT cannot be lower than about
$10^{14}$\,GeV.
\end{abstract}

\pacs{11.10.Kk, 11.15.-q, 12.60.-i
\hspace*{3mm}  
Preprint$^{\#}$: BUHEP-03-06, UTHEP-03-07, OSUHEP-03-2}% 
%PACS, the Physics and Astronomy Classification Scheme.
\maketitle

%%%%%%%%%%%%%%%%%%%%%%%%%%%%%%%%
%\setcounter{section}{0}
%\section{\label{sec:intro}Introduction%
%}%
%--\setcounter{equation}{0}
%\bigskip

%\section{\hspace*{-6mm}.$\!$  Introduction}
%\vspace*{2.5mm}
\noindent %\underline
{\bf 1. Introduction}
\vspace*{2.5mm}

The conventional Higgs mechanism\,\cite{Higgs-64} provides the
simplest way of perturbatively generating gauge boson masses while
ensuring the unitarity of massive gauge boson scattering at high
energies\,\cite{Smith-73,DM-73,Cornwall-74,LQT-77,Veltman-77}.  
Kaluza-Klein (KK)
compactification\,\cite{KK} of extra spatial dimensions, on the
contrary, can geometrically realize vector boson mass generation
without invoking a scalar Higgs particle. 
In this geometric mechanism the
longitudinal components of the massive vector bosons in the
effective four-dimensional (4d) low-energy theory arise from the extra
components of the higher-dimensional gauge field.  The scattering of
massive KK gauge bosons has recently been demonstrated to respect
low-energy unitarity in generic 5d Yang-Mills
theories\,\cite{5DYM:2001a} up to energies inversely proportional 
to the square of the 5d coupling constant.  
In the present Letter, extending this study\,\cite{5DYM:2001a},
we analyze the unitarity of the standard model (SM) 
in $D\,(=4+\delta)$ dimensions,
present a number of general observations, 
and derive their physical consequences.

\vspace*{4.5mm}
\noindent  %\underline
{\bf 2.\,Unitarity\,in\,$D$-Dimensions\,and\,Compactification}
%\vspace*{-1mm}
\vspace*{0.5mm}

We begin with the study of unitarity in the uncompactified
higher-dimensional SM, where the computation of the high energy
scattering amplitudes is simpler than in the compactified theory.
Yang-Mills theories in $4+\delta$ dimensions are not renormalizable.
The gauge coupling constant $\widehat{g}$ has mass dimension
$-\delta/2$, and therefore we expect such a theory can only be an
effective theory valid up to an ultraviolet (UV) cutoff $\Lambda$ of
order ${\widehat{g}}^{-2/\delta}$.  This effective description can
only hold so long as the relevant high energy scattering amplitudes
remain unitary, and we may estimate $\Lambda$ by determining the scale
at which the tree-level scattering amplitudes violate unitarity.

We start by considering the QCD sector of the $D\,(=4+\delta)$ dimensional
SM, in which the gauge bosons and the Higgs doublet propogate in the extra
$\d $ dimensions.  The analysis of $D$-dimensional QCD (D-QCD) is a direct
application of the study of 5d Yang-Mills theory\,\cite{5DYM:2001a}.  The
Yang-Mills symmetry is $SU(k)$ with $k=3$, and the gauge Lagrangian is,
\beq
\ba{ll}
\label{eq:D-QCD}
\widehat{\cal L}_{QCD}^{D} & = \dis
-\f{1}{2}{\rm Tr}(\widehat{G}_{MN}\widehat{G}^{MN}) \,,
\\[3mm]
\widehat{G}_{MN}^a  & 
=\partial_M\Gh^a_N-\partial_N\Gh^a_M+\gsh C^{abc}\Gh^b_M\Gh^c_N \,,
\ea
\eeq
where, as noted above, the $D$-dimensional gauge coupling $\gsh$ has a
negative mass dimension $-\d/2$.  The gauge-fixing and Faddeev-Popov
ghost terms
$\,\widehat{\cal L}_{\rm gf}+ \widehat{\cal L}_{\rm FP}$\, 
can be constructed accordingly.

Consider gluon scattering, $\,\Gh^a_{j_1}\Gh^b_{j_2}\to
\Gh^c_{j_3}\Gh^d_{j_4},\,$ where the index $j\in (1,2,\cdots,D-2)$
denotes the polarization states of gluon field $\Gh^{aM}$. We expect
that scattering amplitude will behave at high-energies as a constant of
${\cal O}(\widehat{g}_s^2)$. As in \cite{5DYM:2001a}, we expect a large
elastic scattering amplitude in the spin-0 and gauge-singlet
two-particle state,
\beq
\label{eq:Dspin0gaugeS}
|\Psi_0\rangle = \dis\f{1}{\sqrt{(D-2)(k^2-1)}}
\sum_{j=1}^{D-2}\sum_{a=1}^{k^2-1} |\Gh^a_j\Gh^a_j\rangle \,.
\eeq
After a lengthy calculation \cite{5DSM:2001}, we derive the scattering
amplitude for this spin-0 and gauge-singlet channel,
\beq
\label{eq:D2to2}
\Th_0[|\Psi_0\rangle\to|\Psi_0\rangle] =
\dis\f{2k\ggsh}{D-2}\[\f{\,4(D-2)\,}{\sin^2\theta}-(D-4)^2\]\!,
\eeq
where $\theta$ is the scattering angle.
The $D$-dimensional $s$-partial wave amplitude\,\cite{Soldate}
is thus deduced as
\beq
\label{eq:a0-GGD}
\ba{ll}
\widehat{a}_{00}^{~} & =\, \dis
\f{\(\!\sqrt{s}\,\)^{\d}}{2(16\pi)^{1+\f{\d}{2}}\,\Gamma\(1+\f{\d}{2}\)}
\int^\pi_0 \! d\theta\,(\sin\theta)^{1+\d} \,\Th_0
\\[4.5mm]
& = \dis 
\f{2k\sqrt{\pi}\[4(2+\d)-\f{\d^3}{1+\d}\]}
  {(16\pi)^{1+\d/2}\d(2+\d)\Gamma\(\f{1+\d}{2}\)}
\(\ggsh\,s^{\f{\d}{2}}_{~}\)~.
\ea
\eeq

With the unitarity condition 
\,$|\RE\,\widehat{a}_{00}^{~}|<{2!}/{2}$\, 
(where the factor $2!$ is due to the identical particles 
in the final state), we arrive at
\beq
\label{eq:UB-DQCD}
\sqrt{s} < \dis
\[\f{\dis(16\pi)^{1+\f{\d}{2}}\d(2+\d)\,\Gamma\!\(\f{1+\d}{2}\)}
    {\dis2k\sqrt{\pi}
     \left|4(2+\d)-\f{\d^3}{1+\d}\right|} \f{1}{\,\ggsh\,}
\]^{1/\d},
\eeq
which shows that tree-level unitarity violation in such a
$D$-dimensional theory indeed occurs at an energy of order its
intrinsic ultraviolet (UV) scale ${\cal O}(\widehat{g}_s^{-{2}/{\d}})$.

For the gluon scattering being described by this effective
higher-dimensional gauge theory, the scattering energy $\sqrt{s}$
cannot exceed the cutoff $\cut$ of the effective theory,
$ \sqrt{s}< \cut$\,.\,
Therefore, for the highest possible scattering energy,
the scale $\cut$ is also bounded 
by the right hand side of Eq.\,(\ref{eq:UB-DQCD}), 
which results in a constraint on $\ggsh$\,,
\beq
\label{eq:UB-Dgs2}
\ggsh ~<~ \dis
\f{\dis ~(16\pi)^{1+\f{\d}{2}}\d(2+\d)\Gamma\!\(\f{1+\d}{2}\)}
    {\dis 2k\sqrt{\pi}\left|4(2+\d)-\f{\d^3}{1+\d}\right|} 
  \f{1}{\,~\cut^\d\,} .
\eeq

The uncompactified higher-dimensional SM by itself is, of course, of
no direct phenomenological relevance.  
Rather, we would like to consider the
situation when the $\delta$ extra spatial dimensions are compact. In
this case, each extra dimension is manifested as Kaluza-Klein towers
of the given fields, with mass spectra characterized by $|n|/R$, where
$1/R$ is the compactification scale and $n$ the KK-level. The details
of the spectra and interactions of these modes depend on 
the extra-dimensional boundary conditions imposed 
at the scale $1/R$.  In the next section, we will
explicitly investigate the unitarity of the 5d SM compactified on
$S^1/\ZZ_2$.  However, we expect the essential properties of unitarity
-- which arise from the high-energy/short-distance behavior 
($E\gg 1/R$) of the higher-dimensional gauge theory 
-- should be insensitive to the
details of the compactification.  In particular, we expect that
Eq.\,(\ref{eq:UB-Dgs2}) should hold at least approximately after the
extra dimensions are compactified.

In the following, we investigate the consequences of the bound
Eq.\,(\ref{eq:UB-Dgs2}) as applied to a toroidal compactification of the
$\d$ extra dimensions on $T^\d = (S^1/\mathbb{Z}_2)^\d$, with a common
radius $R$.  (For $\d \geqq 2$, the current analysis may be readily
extended to the case of asymmetric compactification\,\cite{LN} or with 
nontrivial shape moduli\,\cite{shape}, with which some of the constraints 
could be relaxed\,\cite{5DSM:2001}.)

The $D$-dimensional gauge coupling $\ggsh$ is connected to the
dimensionless coupling $g_s^2$ in the compactified effective 4d KK
theory via, $\,\ggsh = V_\d g_s^2\,$ with the $D$-dimensional volume
$V_\d = (\pi R)^\d$.  As described previously, we expect the coupling
$\,\ggsh$ to be bounded for a given value of the cutoff $\Lambda$. The
effective 4d KK theory can only contain modes of a mass
below the cutoff $\Lambda$. Therefore, the highest
KK level in this compactification
is fixed by \,$\NKK =\cut R$\,.\, We can thus convert
Eq.\,(\ref{eq:UB-Dgs2}) into a constraint on $\NKK$,
\beq
\label{eq:UB-DNKK}
\NKK < \dis
\[\f{\dis 4^{2+\d}\pi^{\f{1-\d}{2}}\d\,(2+\d)\,
     \Gamma\!\(\f{1+\d}{2}\)}
    {\dis 2k\left|4(2+\d)-\f{\d^3}{1+\d}\right|} 
    \f{1}{~g_s^2~}
\]^{1/\d},
\eeq
which, as written, applies to any compactified $SU(k)$
Yang-Mills theory.

Eq.\,(\ref{eq:UB-DNKK}) shows that for a given value of the
four-dimensional gauge coupling, the highest KK level in the
effective 4d theory is bounded from above. 
The emergence of this bound reflects the fact that, 
as shown in Ref.\,\cite{5DYM:2001a}, the bad high-energy behavior 
of the underlying higher-dimensional gauge theory
manifests itself through the appearance of the myriad KK excitations 
which couple together to enhance the scattering amplitude
and thus speed up unitarity saturation.

To illustrate these features, we first apply the bound
(\ref{eq:UB-DNKK}) to the QCD sector.  Taking a sample value of
$\alpha_s\simeq 0.1$ at ${\cal O}({\rm TeV})$ scale, we derive the following
bounds from Eq.\,(\ref{eq:UB-DNKK}),
\beq
\label{eq:UB-DNKKqcd}
\NKK \leqq (2,\,3,\,3,\,4,\,4,\,6,\,4)\,,
\eeq
for $\d = D-4 =(1,\,2,\,3,\,4,\,5,\,6,\,7)$, respectively.  
It is also straightforward 
to apply the condition (\ref{eq:UB-DNKK}) to the
scattering of weak gauge bosons in the $SU(2)_W\otimes U(1)_Y$
electroweak (EW) gauge sector of the $D$-dimensional SM. Ignoring
the small $W_3$-$B$ mixing and 
setting $SU(k)=SU(2)_W$, we deduce the following
bounds on $\NKK$, for $\d =(1,\,2,\,3,\,4,\,5,\,6,\,7)$,
\beq
\label{eq:UB-DNKKew}
\NKK \leqq (9,\,6,\,6,\,5,\,6,\,8,\,6)\,,
\eeq
where we have replaced $g_s^{-2}$ by $g^{-2} = (v/2m_w)^2$ in
(\ref{eq:UB-DNKK}), with $v$ the Higgs vacuum expectation value (VEV)
and $m_w$ the mass of weak gauge bosons.  For $\d \leqq 3$, this is
significantly weaker than (\ref{eq:UB-DNKKqcd}) from the QCD sector.
The non-monotonic behavior of $\NKK$ bounds (\ref{eq:UB-DNKKqcd}) and
(\ref{eq:UB-DNKKew}) as a function of $\d$ is because in the 
right-hand side of
Eq.\,(\ref{eq:UB-DNKK}) the denominator has a maximum at $\d\simeq
2.5$ and flips sign at $\d\simeq 6.2$.

Next, we analyze the scalar Higgs 
sector of the $D$-dimensional SM.
The Higgs Lagrangian is
\beq
\label{eq:DLagH}
\widehat{\cal L}_h^{D} =
D_M^{~}\Phih^\dag D^M\Phih 
- \[-\mu^2\Phih^\dag\Phih + \lah (\Phih^\dag\Phih)^2
     \] ,
\eeq
where $ D_M^{~}\Phih = (\partial_M-\f{i}{2}{\gh}\tau^a\Wh^a_M
-\f{i}{2}{\gh'}\Bh_M)\Phih $,\, and the $D$-dimensional Higgs (gauge)
coupling $\lah$ $(\gh,\gh' )$ has a mass dimension of $-\d$
$(-{\d}/{2})$.  Due to $-\mu^2 <0$, the Higgs doublet $\Phih$ develops
a VEV so that $\Phih =(-i\pih^+,\, (\vd +\hh^0+i\pih^0)/\sq )^T$,\,
and the neutral Higgs boson $\hh^0$ acquires its mass at tree-level
\beq
\label{eq:mHD}
\widehat{m}_H^{~} = \dis\sqrt{\,2\,\lah\,\vd^2\,} \,.
\eeq

Consider the $D$-dimensional
scalar-scalar scattering, $|\phih^a\phih^b\rangle \to
|\phih^c\phih^d\rangle $, via the neutral channels
$|\phih^a\phih^b\rangle, |\phih^c\phih^d\rangle =
\f{1}{\sq}|\hh^0\hh^0\rangle,\,
\f{1}{\sq}|\pih^0\pih^0\rangle,\, |\pih^+\pih^-\rangle,\,
|\hh^0\pih^0\rangle$, in analogy with the customary analysis
in the usual 4d SM Higgs sector\,\cite{LQT-77}.
Their scattering amplitudes form a $4\times 4$
matrix and, for $s\gg\widehat{m}_H^2$, approach a constant matrix
which arises from the four-scalar contact interaction.  
(The channel $|\hh^0\pih^0\rangle$ actually decouples from the 
other three.)
The eigenvalues of this constant matrix are readily worked out
as  $\,-\lah \cdot \(6,\,2,\,2,\,2\)\,$ so that
the maximal eigenvalue amplitude is
$\,\widehat{\TT}_{\max} = -6\lah \,$.\,
Thus, using the first formula in Eq.\,(\ref{eq:a0-GGD}),
we derive the $D$-dimensional $s$-partial wave amplitude,
\beq
\label{eq:a0-D}
\ba{l}
\hspace*{-3mm}
\widehat{a}_0^{~} 
%=\,\dis\f{\(\!\sqrt{s}\,\)^{\d}}{2(16\pi)^{1+\f{\d}{2}}\,
%\Gamma\(1+\f{\d}{2}\)}\int^\pi_0 \! d\theta\,
%(\sin\theta)^{1+\d} \,\widehat{\TT}_{\max} 
\,=\, \dis -C_0\, \(\!\sqrt{s}\,\lah^{\f{1}{\d}}_{~}\)^{\d}, 
~~~
C_0\equiv\dis \f{ 3\sqrt{\pi}}
          {~(16\pi)^{1+\f{\d}{2}}\,\Gamma\(\f{3+\d}{2}\)~} .
\ea
\eeq
The relevant unitarity condition is
\,$|\RE\,\widehat{a}_0^{~}|<\f{1}{2}$\,,\,
where the possible identical particle factors 
are included\,\cite{LQT-77} 
in the normalization of in/out state
$|\phih^a\phih^b\rangle$ or 
$|\phih^c\phih^d\rangle$ mentioned above.
So, we deduce, 
%from Eq.\,(\ref{eq:a0-D}), a bound
%
\beq
\label{eq:UB-Ds}
\sqrt{s} \,~<~ \dis \(2\,C_0\,\lah\)^{-\f{1}{\d}} \,,
\eeq
which again shows that the unitarity violation in such a
$D$-dimensional theory indeed occurs at the intrinsic UV
scale, of  ${\cal O}(\lah^{-{1}/{\d}})$ in this case. 
Similar to Eq.\,(\ref{eq:UB-Dgs2}), 
we may interpret Eq.\,(\ref{eq:UB-Ds})
as an upper bound on the cutoff $\cut$, and translate
this into a condition on \,$\lah$\,,
\beq
\label{eq:UB-DHcoup}
\lah \,~<~ \dis\f{1}{~2\,C_0(\sqrt{s}\,)^\d_{\max} ~}
             = \f{1}{~2\,C_0\, \cut^{\d}~} \,.
\eeq

In parallel with our analysis of the gauge sector, we interpret
(\ref{eq:UB-DHcoup}) in the context of compactification on
$T^\d=(S^1/\ZZ_2)^\d$ and examine the consequences for the effective 4d
KK-theory.  The $D$-dimensional coupling $\lah$ is related to the
dimensionless coupling $\la$ in the 4d KK-theory via, $\,\lah =V_\d\la
=(\pi R)^\d\la\,,$\, and Eq.\,(\ref{eq:UB-DHcoup}) can be rewritten as
\beq
\label{eq:UB-4dHcoup}
\la \,~<~ \dis 
          \f{1}{~2\,C_0 V_\d \cut^\d~}
   \simeq \f{1}{~2\,C_0 \pi^\d \(N_{\rm KK}\)^\d~}\,,
\eeq
where \,$N_{\rm KK}\simeq \cut R$\, 
represents the highest KK levels 
associated with each compactified extra dimension in the
low-energy effective theory.  Here, for simplicity, we have ignored
the small $\,{\cal O}(m_w^2 R^2)\,$ correction to the KK mass spectrum in
the EW sector.

The 4d KK theory contains a zero-mode Higgs doublet $\Phi_0 =
(-i\pi_0^+,\, (v + h_0^0+i\pi_0^0)/\sq \,)^T$,\, and its KK
excitations $\Phi_{\rm{\bf n}} = (-i\pi_{\nb}^+,\, (v +
H_{\nb}^0+i\pi_{\nb}^0)/\sq \,)^T$, where the VEV of the 4d Higgs
doublet $\Phi_0$, $v=(\sq G_F)^{-1/2}\simeq 246$\,GeV, is related to
that of $\Phih$ via \,$\vd \,=\, v/\sqrt{V_\d}$\,.\, The corresponding
mass of the zero-mode neutral Higgs boson $h_0^0$ is then given by
\beq
\label{eq:mH-4-D}
\mH \,=\, \dis\sqrt{\,2\la v^2\,} ~=~ 
          \sqrt{\,2\lah\, \vd^2\,} 
    \,=\, \widehat{m}_H^{~} \,,
\eeq
which is unchanged under the compactification.
Hence, we can deduce, from (\ref{eq:UB-4dHcoup}), the unitarity bound
on the physical mass of the 4d SM Higgs boson $h_0^0$\,,
\beq
\label{eq:UB-mH-D4}
\mH \,~<~ 
\dis \f{v}{\,\sqrt{C_0 V_\d \cut^{\d}_{~}\,}~}
\simeq \f{v}{\,\sqrt{C_0 \pi^\d \(N_{\rm KK}\)^\d~}\,}\,,
\eeq
where for $\d = (1,2,3,4,5,6,7)$,
$$
1/\sqrt{\,C_0\,\pi^\d\,}
\simeq (4.6,\, 8.0,\, 14.7,\, 28.5,\, 57.6,\, 121,\, 260)\,.
$$
Alternatively, for a given $\mH$, 
\beq
\label{eq:UB-NKKH}
\NKK  \,~<~ \dis 
           \(\f{v^2}{\,C_0 \pi^\d \,m_H^2\,}\)^{\f{1}{\d}}\,.
\eeq

A few comments are in order. We note that Eq.\,(\ref{eq:UB-NKKH})
shares similar features to the gauge KK bound (\ref{eq:UB-DNKK}),
except that in the Higgs sector the coupling $\la$ (or mass $\mH$) is
not fixed by observation.  If we impose the existing direct Higgs
search limit at LEP-2, $\mH > 114.3$\,GeV (95\%\,C.L.), we can deduce,
from (\ref{eq:UB-NKKH}),
\beq
\label{eq:NKK-mH115}
\NKK \leqq  (98,\,17,\,10,\,7,\,6,\,6,\,6)\,,
\eeq
for $\d = (1,2,3,4,5,6,7)$.\, 
The limits (\ref{eq:NKK-mH115}) are
significantly weaker than the bounds
(\ref{eq:UB-DNKKqcd})-(\ref{eq:UB-DNKKew}) from the gauge KK sector,
especially for $\,\d \leqq 3$\,.\, The bound (\ref{eq:UB-mH-D4}) on the
Higgs-boson mass is listed in Table\,I for various values of
$\delta$ and $\NKK$. The entries marked by \as\, 
are excluded by direct Higgs
searches at LEP-2, cf. Eq.\,(\ref{eq:NKK-mH115}).  
We see that $\NKK$ is more severely
bounded by the unitarity of the gauge boson scattering, 
especially for gluons as long as they propagate
in the bulk [cf. (\ref{eq:UB-DNKKqcd})].  
Imposing the stronger bound (\ref{eq:UB-DNKKqcd}), 
we can examine how Eq.\,(\ref{eq:UB-mH-D4}) would
constrain $\mH$.  
The corresponding upper limits are displayed in the entries
of Table\,I marked by \sst\,.
%
%
%\vspace*{-3mm}
\begin{table}[H]
\label{Tab:Tab1}
\vspace*{-3mm}
\caption{Estimated limits on $\mH$ (in GeV)
from Eq.\,(\ref{eq:UB-mH-D4}) 
as \,$\d$\, and \,$\NKK$\, vary. 
The entries marked by \as\, are already
excluded by LEP-2 searches, while only those marked
by \sst\, are allowed after imposing the $\NKK$ bound
(\ref{eq:UB-DNKKqcd}).
}
\vspace*{-5mm}
\begin{center}
\begin{tabular}{r||ccccccccc}
\hline\hline
&&&&&&&&&\\[-2.5mm]
$\NKK =$~\, & 3 & 4  & 5  & 6 &  8  &  10  &  12  &  15  &  20 \,
\\ [1.5mm]
\hline %\hline
&&&&&&&&&\\[-2.5mm]
~~$\d =1$~
& 656 & 568 & 508 & 464  & 402 & 359 & 328 & 293 & 254  
\\ [1.5mm]
%\hline
%&&&\\[-2.5mm]
~~$ =2$~
& 656\sst & 492 & 394 & 328 & 246 & 197 & 164 & 131 & 98\as  
\\ [1.5mm]
~~$ =3$~
& 698\sst & 453 & 324 & 247 & 160 & 115 &  87\as & 62\as &  41\as  
\\ [1.5mm]
~~$ =4$~ 
& 780\sst & 438\sst & 281 & 195 & 110\as  & 70\as &  49\as & 31\as &  18\as 
\\ [1.5mm]
~~$ =5$~ 
& 909\sst & 443\sst & 254 & 161 & 78\as   & 45\as &  28\as & 16\as &  8\as 
\\ [1.5mm]
~~$ =6$~
& 1098\sst & 463\sst & 237\sst & 137\sst & 58\as  & 30\as &  17\as & 9\as &  4\as
\\ [1.5mm]
~~$ =7$~
& 1368\sst & 500\sst & 229 & 121 & 44\as  & 20\as &  11\as & 5\as &  2\as
\\ [1.5mm]
\hline\hline
\end{tabular}
\end{center}
\vspace*{-5mm}
\end{table}
%
%\vspace*{-3mm}
%

Finally, we stress that the strong bounds we have derived for either 
the coupling constants $(g^2,\,\lambda)$
or the Higgs mass $\mH$ are due to the {\it
large extra dimensional volume} $V_\d\sim R^\d$ which appears in
relating the $D$-dimensional and four-dimensional physics.
Equivalently, following\,\cite{5DYM:2001a}, the strong bounds arise because
of the appearance of {\it many KK excitations} in the compactified
theory which couple together to enhance the scattering amplitude and
speed up unitarity saturation.  Therefore, it is clear that if any
field is restricted on a brane, the corresponding unitarity bound on
its couplings would reduce back to that of the customary 4d SM.

In the scenario with universal extra dimensions\,\cite{UED}, all SM
fields propagate in the extra dimensions and are thus subject to the
unitarity limits discussed above. In particular, these results suggest
that the ``self-broken'' standard model\,\cite{Arkani-Hamed:2000hv}
would not be realized consistently. For a given $\NKK$ and under the
compactification $(S^1/\ZZ_2)^\d$, the total number of states
consistent with the unitarity constraint is approximately given by
\beq
n(\NKK ) \,\approx\, 
\f{\pi^{\delta/2}}{\,\Gamma\left(1+{\f{\delta}{2}}\right)\,}
\left(\f{\NKK}{2}\right)^\delta~.
\eeq
For $D=6$, our constraints in (\ref{eq:UB-DNKKqcd})
yield a total number of KK modes \,$n \leqq
7$,\, and for $D=8$ we find \,$n \leqq 79$.\, Neither appears likely
to yield a sufficiently light top quark mass at the observed value,
for a wide range of compactification
scales\,\cite{Arkani-Hamed:2000hv}.

For theories with extra dimensional perturbative gauge
unification\,\cite{powerlaw}, all gauge fields live in the bulk and
again these bounds apply.  We will return to implications for GUTs in
Sec.\,4.  Finally, we note that our analysis may be
extended to bounds on gauge KK scattering and graviton KK
scattering\,\cite{5DSM:2001} in a warped 5d SM \`{a} la
Randall-Sundrum (RS1)\,\cite{RS1}.

\vspace*{4.5mm}
\noindent   %\underline{
{\bf 3.\,KK\,Theory:\,$E$-Cancellations\,and\,Unitarity\,Limits
}
%\vspace*{-1mm}

We now turn to an explicit analysis of the effective 4d KK theory
arising from compactifying a 5d SM on $S^1/\ZZ_2$.  The Lagrangian of
this 4d KK theory will contain interactions involving purely physical
fields\,\cite{4DKK-LU,Hill} and interactions involving additional
geometric and ordinary would-be Goldstone
fields\,\cite{5DYM:2001a,5DRxi2}.  In Ref.\,\cite{5DYM:2001a}, the
scattering of the longitudinal components of the KK excitations of
Yang-Mills fields was systematically computed.  Here we will further
compute all amplitudes involving transversely polarized gauge KK
fields, and in the case of the EW gauge bosons, include the effect of
EW symmetry breaking of the SM.

We begin by considering gluon KK scattering $G^{a,n}_{j_1}G^{b,n}_{j_2}\to
G^{c,\l}_{j_3}G^{d,\l}_{j_4}$, where $j\in (+,0,-)$ denotes the three
helicity states and $(n,\l)$ the KK levels. We may define a spin-0,
gauge-singlet state,
\vspace*{-2mm}
\beq
\label{eq:KKspin0gaugeS}
|\Psi_0^n\rangle = \dis\f{1}{\sqrt{3(k^2-1)}}
\sum_{j=-1}^{+1}\sum_{a=1}^{k^2-1} 
\left| G^{a,n}_j G^{a,n}_j\right\rangle \,,
\eeq
where $k=3$ for QCD $SU(3)_c$. 
The corresponding $|\Psi_0^n\rangle \to |\Psi_0^{\l}\rangle$ 
scattering channel consists of 9 helicity amplitudes, but
only 4 of them are independent under the discrete 
(\tx{P}, \tx{C}, \tx{T}) symmetries. We arrive at
\beq
\label{eq:TKK-00}
\ba{l}
\hspace*{-2mm} 
\T_0 \[|\Psi_0^n\rangle\to|\Psi_0^{\l}\rangle\]
~=~  
%\!\!\!
\dis\f{1}{3(k^2\!-\!1)} \!\!\!
\sum_{j,j'=-1}^{+1}\sum_{a,c=1}^{k^2\!-\!1}  \!\!
\\[4.7mm] 
\hspace*{7mm}
\left\{\T^{aa,cc}_{00,00} +2\T^{aa,cc}_{++,++} 
%\\[4.5mm]
% & \hspace*{28mm}  
+2\T^{aa,cc}_{++,--}+4\T^{aa,cc}_{00,++}\right\}.
\ea
\eeq
A systematic calculation shows 
that the amplitudes $\T^{aa,cc}_{00,++}$ and
$\T^{aa,cc}_{++,++}$ vanish to ${\cal O}(g_s^2E^0)$, while the amplitudes
$\T^{aa,cc}_{00,00}$ and $\T^{aa,cc}_{++,--}$ both have nonzero
${\cal O}(g_s^2E^0)$ contributions. We have
verified the nontrivial energy-cancellations at 
${\cal O}(E^{4,2})$ [${\cal O}(E^{2})$]
for the amplitudes $\T^{ab,cd}_{00,00}$ [$\T^{ab,cd}_{00,++}$]
involving four [two] external longitudinal KK gluon states, 
and the consistency with the
Kaluza-Klein Equivalence Theorem (KK-ET)\,\cite{5DYM:2001a}.  
We then compute the $s$-partial wave amplitude 
for (\ref{eq:TKK-00}) with $n\neq \l$,
\beq
\label{eq:a00-KK}
a_{00}^{~}=\dis\f{\,kg_s^2\,}{24\pi}
           \[-1+6\ln\f{N_s^2}{|n^2-\l^2|}\],
\eeq
where $\,N_s\equiv \sqrt{s}\,R \leqq \NKK$\,.\,
This explicitly shows that due to the 
exact $E$-cancellations, the partial wave amplitude
indeed behaves as constant at the leading order.

To maximize the scattering amplitude, 
we define a normalized state consisting
of KK-levels up to $N_0$, %i.e., 
\beq
|\Omega\rangle = \dis
\f{1}{\sqrt{N_0}}\sum_{n=1}^{N_0} |\Psi_0^n\rangle ,
\eeq
where the kinematics of
$2\to 2$ scattering requires $N_0 < N_s/2$.  For
$(N_0)_{\max}=(N_s)_{\max}/2=\NKK/2$, we deduce the maximal $s$-wave
amplitude for $|\Omega\rangle\to |\Omega\rangle$, to leading 
order in $\NKK$,
\beq
\label{eq:a00F}
a_{00}^{~}[\Omega]=\dis\f{kg_s^2}{8\pi}\!
\[-\f{\NKK}{6}+\f{4}{\NKK}\!\sum_{n\neq\l=1}^{\NKK/2}
\!\ln\f{\NKK^2}{|n^2-\l^2|}\]\!.
\eeq
From the unitarity condition 
$|a_{00}^{~}|<\sqrt{2!}/2$ (where the $2!$ arises
from identical particles in the final state\,\cite{5DSM:2001}),
we derive the following numerical bound for the QCD sector,
\beq
\label{eq:NKK-4dqcd}
\NKK ~ \leqq ~ 4\,.
\eeq

We have performed a similar analysis 
for KK scattering in the EW gauge sector, 
where a small mixing\,\cite{5DRxi2} arises 
between the geometric KK Goldstone bosons $W_n^{a5}$ 
and the KK excitation modes $\pi_n^{a}$ of
ordinary Goldstone bosons in the Higgs doublet. 
This mixing is described  
by a mixing angle $\,\sin\theta_n = m_{v}/M_n \ll 1$, 
and $m_v(=m_{w,z})$ is
the mass of the zero-mode gauge bosons $(W^\pm_0,\,Z^0_0)$.\, 
As a result, the ``eaten'' KK Goldstone field is
\,$\widetilde{\pi}^a_n =
\cos\theta_n\,V_n^{a,5}+\sin\theta_n\,\pi_n^a$,
($V^{a,5}_n=W^{\pm,5}_n,Z^{0,5}_n$), and the gauge KK modes
$V_n^{a,\mu}\, (=W^{\pm,\mu}_n,\,Z^{0,\mu}_n)$ have mass $M_n =
\sqrt{(n/R)^2+m_v^2}\simeq n/R$,\, (\,$m_v \ll n/R$).  There are three
types of physical KK Higgs states $(H_n^{\pm},\,P_n^0)$ which are just
orthogonal to the ``eaten'' KK Goldstone fields
$(\widetilde{\pi}^\pm_n,\,\widetilde{\pi}^0_n)$\,.

From direct calculation\,\cite{5DSM:2001}, 
we find that this small mixing
causes extra $E^2$ contributions to individual contribtuions to the
scattering amplitude with four external longitudinal gauge KK states,
but they exactly cancel to ${\cal O}(E^0)$ for each process.  This is
consistent with the $E$-counting for the corresponding KK Goldstone
amplitude based on the KK-ET\,\cite{5DYM:2001a} which, unlike the
conventional ET for the 4d
SM\,\cite{Cornwall-74,LQT-77,mike,He-ET1,He-ET2}, involves the
geometric Higgs mechanism from compactification.  Analogous to our
analysis in the QCD sector and ignoring the tiny constant terms
suppressed by $m_w^2R^2$ or $m_w^4R^4$, we derive the unitarity bound
on $\NKK$ from a coupled channel analysis for the $2\to 2$ EW gauge KK
scattering,
\beq
\label{eq:NKK-4dew}
\NKK ~\leqq ~ 11 \,,
\eeq
where we have replaced $g_s^2$ in Eq.\,(\ref{eq:a00F}) 
by the weak gauge coupling $g^2=(2m_w/v)^2$.

It is interesting to compare the bounds (\ref{eq:NKK-4dqcd}) and
(\ref{eq:NKK-4dew}) with those estimated from the uncompactified
$D$-dimensional scattering analysis for $D=5$.  We see that
(\ref{eq:NKK-4dew}) agrees with (\ref{eq:UB-DNKKew}) quite well 
where the $\NKK$ upper limits are about $9-11$, 
while (\ref{eq:NKK-4dqcd}) agrees
with (\ref{eq:UB-DNKKqcd}) up to a factor of $2$ where the $\NKK$ is
constrained to be no higher than the range of $2-4$.  This is as
expected, however, since for very low values of $\NKK $ the kinematic
effects due to the finite KK masses (which are absent in the
uncompactified analysis) would become more important.
Also, the subleading terms ignored in the 4d amplitude (\ref{eq:a00F})
are suppressed by a factor $1/\NKK$ relative to the leading terms and
imply a larger uncertainty for very low values of $\NKK$. Hence, we
see that the two independent analyses are consistent with each other,
and they provide consistent estimates for the
unitarity bounds.

Next, we perform a coupled channel analysis 
for the $2\to 2$ Higgs KK scattering and derive
the $\mH$ bounds in the effective 4d KK theory.
There are four types of processes (and their
crossing channels) which appear relevant
to this coupled channel analysis:
(i)   $\HB_n \HB_n       \to \HB_k \HB_k$,\,
(ii)  $h_0^{~}\HB_{2k}   \to \HB_k \HB_k$,\,
(iii) $h_0^{~}h_0^{~} \to \HB_k \HB_k$,\,
(iv)  $V_{0L}^a V_{0L}^a \to \HB_k \HB_k$,\,
where $\HB_n \in (H_n^0,\,P_n^0,\,H_n^\pm)$ represents
three types of physical Higgs KK states and 
$V_{0L}^a\in (W_{0L}^\pm,\,Z_{0L}^0)$ denotes the zero-modes
of the longitudinal weak gauge bosons.
As will be clear shortly, we find that the only important processes
for our coupled channel analysis are type-(i) which involves
only KK scalars (without zero-mode) 
for the in/out states of the $2\to 2$ scattering.

For the type-(i) channels, we will consider the scattering
$|\HB_n\HB_n\rangle \to |\HB_k\HB_k\rangle$,\,
via electrically neutral KK channels
$|\HB_n\HB_n\rangle,\,|\HB_k\HB_k\rangle
=\f{1}{\sqrt{2}}|H_{\l}^0 H_{\l}^0\rangle,\,$ \linebreak
$\f{1}{\sqrt{2}}|P_{\l}^0 P_{\l}^0\rangle,\, 
 |H_{\l}^+ H_{\l}^-\rangle,\,
 |H_{\l}^0 P_{\l}^0\rangle$  (${\l} =n,k$) 
in analogy with the customary analysis of the 4d SM\,\cite{LQT-77}.
Again, the channel $|H_{\l}^0 P_{\l}^0\rangle$ decouples from the
other three channels and
their scattering amplitudes form a $4\times 4$
matrix which approaches constant
for $s\gg M_{\HB}^2$.
We then derive the eigenvalues of this matrix as,
$\,-\la \cdot \(6,\,2,\,2,\,2\)\,$ for
$n\neq k$, 
and $\,-(3\la /2)\cdot \(6,\,2,\,2,\,2\)\,$ 
for $n=k$. 
Thus, the maximal eigenvalue amplitudes are
\beq
\label{eq:Amp-4Hn}
%\ba{lcl}
\dis 
{\T}_{\max}[nn,kk] = -6\la \,,~~~\,
%\\[2.5mm]
\dis 
{\T}_{\max}[nn,nn] = -9\la \,,
%\ea
\eeq
where \,$n\neq k$\,.\,
Defining a normalized state consisting of $N_0$ pairs of
KK states, 
\beq
\label{eq:coupS}
|S\rangle ~=~ \dis\f{1}{\sqrt{N_0}}\sum_{n=1}^{N_0} 
                  |nn\rangle \,,
\eeq
we deduce the $s$-wave amplitude for 
$|S\rangle\to |S\rangle$, at the leading order of 
\,$N_0$\,,
\beqa
\label{eq:TS-4H0}
a_0^{~}[S] &=& -\dis\f{3N_0}{16\pi}\(\f{\mH}{v}\)^2 \,,
\eeqa 
where the inelastic channels $\,nn\to kk$\, 
($n\neq k$) dominate while channels $\,nn\to nn$\,
are only of ${\cal O}((N_0)^0)$.
From the unitarity condition \,$|\RE\,a_0^{~}|< 1/2$\,
and noting the kinematic requirement
\,$N_0\leqq \NKK/2 \simeq \cut R/2$,\, 
we deduce the $\NKK$ limit,
\beq
\label{eq:UB-4d-HNKK}
\NKK ~<~ \dis
\f{\,16\pi\,}{3}\(\f{v}{\,\mH\,}\)^2 \,,
\eeq
or, the bound on the Higgs mass,
\beq
\label{eq:UB-4d-mH}
\mH ~<~ \(\f{\,16\pi\,}{3}\)^{\f{1}{2}}
        \f{v}{\,\sqrt{\NKK}\,} \,.
\eeq
With these we can constrain $\NKK$ by imposing the
LEP-2 Higgs search limit $\mH < 114.3$\,GeV,
\beqa
\label{eq:UB-4d-HNKKx}
\NKK &\leqq& 77 \,. 
\eeqa
Comparing our estimated bound (20) for $D=5$
with the above limit, we see that the difference is only
about $21\%$. 
Using the condition (\ref{eq:UB-4d-mH}), 
we further derive the Higgs mass limits,
\beq
\label{eq:UB-4d-mHx}
\hspace*{-2mm}
\mH  <  (581,\,503,\,450,\,411,\,356,\,318,\,
         291,\,260,\,225)\,{\rm GeV},
\eeq
for the inputs $\NKK =(3,4,5,6,8,10,12,15,20)$, respectively. Again,
we notice that the estimated bounds in Table\,I ($\d=1$) are in
reasonable agreement.

Finally, we comment that for type-(ii), -(iii) and -(iv) processes 
we could define similar
normalized in/out state to Eq.\,(\ref{eq:coupS}), but it is
readily seen that the corresponding scattering amplitudes only
have leading contributions at 
${\cal O}(1)$, ${\cal O}(\sqrt{N_0})$ and  
${\cal O}(\sqrt{N_0})$, respectively.
The type-(i) channels with $n=k$ also have amplitudes of 
${\cal O}(1)$.
So, the bounds from these other channels
are too weak to be useful in comparison with 
that of type-(i) with $n\neq k$.

\vspace*{4.5mm}
\noindent   %\underline
{\bf 4. Gauge Unification and Unitarity Constraints}
\vspace*{2.5mm}

In higher dimensional Grand Unified Theories (GUTs), the KK states
will contribute to the running gauge coupling constants.  For
sufficiently many KK states, these contributions mimic the power-law
running expected in a higher dimensional theory and it has been
proposed that this could substantially lower the GUT scale in these
theories\,\cite{powerlaw}.  The unitarity bounds we derived from the
gauge KK scattering in the previous sections can apply to any higher
dimensional GUT (with/without supersymmetry) whose low energy theory
contains the SM gauge bosons and their KK excitations as part of the
spectrum. We will show that such unitarity constraints severely
restrict the number of KK states which can consistently accelerate
perturbative gauge-coupling unification.  This bound prevents the
unification scale in 5d minimal SUSY GUT from being lower than
about $10^{14}$\,GeV.

In the $\overline{\rm MS}$ scheme,
the running gauge coupling may be expressed 
as\,\cite{powerlaw,Hill-GUT}
\beq
\label{eq:gGUT}
%\ba{ll}
\dis
\hspace*{-1.5mm}
\ainv_j\!(\mu) ~=~ \dis\ainv_{zj}
 -\f{b_j}{2\pi}\ln\!\f{\mu}{m_z}
 -\f{\widetilde{b}_j}{2\pi}F(\d,n(\mu))
  + \k_j \,,
%\ea
\eeq
where $\a_{zj}\equiv \a_j(m_z)$,\,
$\k_j$ represents corrections from the higher
loop-levels and possible higher
dimensional operators suppressed by GUT scale $\MG$ \cite{Hill-GUT}.
The $ F(\d,n(\mu))$ term arises 
from the one-loop KK contributions, 
\beq
\label{eq:F}
F(\d,n(\mu)) ~=~ \dis
\sum_{n=1}^{n(\mu)} {\cal D}_n\ln\f{\mu}{M_n}\,,
\eeq
where $M_n$ is the mass of relevant KK excitations at level $n$, 
$n(\mu)$ is defined by 
\,$M_{n(\mu)} < \mu < M_{n(\mu)+1}$,\, and
$\,1\leqq n(\mu) \leqq \NKK $.
The ${\cal D}_n$ denotes the degeneracy at the KK level $n$.\,

For $\d=1$ with compactification on $S^1/\ZZ_2$, we have
${\cal D}_n=1$ and
\beq
F(1,\NKK)= \NKK\ln\NKK - \ln(\NKK!) \,,
\eeq
where we set $\NKK = R M_G$.\, Unification at the GUT scale
imposes the conditions
\,$\a_1(\MG) = \a_2(\MG) =\a_3(\MG)$,\, 
where we have used the usual GUT normalization,
\,$\alpha_1=(5/3){g'}^2/(4\pi)$\,.
From this, we arrive at
\beq
\label{eq:MGcond}
%\ba{l}
\hspace*{-1.5mm}
\MG=m_z\dis\exp\left\{
\!\f{2\pi}{\D b_{ij}}\!\!\[
\D\alpha_{zij}^{-1}-\f{\D\bt_{ij}}{2\pi}F(\d,\NKK)+\D_{ij}
\]\!
\right\},
%\ea
\eeq
where 
\,$\D\alpha_{zij}^{-1}=\alpha_{zi}^{-1}-\alpha_{zj}^{-1}$,\,
  $\D b_{ij} = b_i - b_j $,\,
  $\D\bt_{ij} = \bt_i - \bt_j $,\,
  $\D_{ij} = \kappa_i - \kappa_j
         = {\cal O}(10^{-2})$,\,
and \,$i<j =1,2,3$.\,
Eq.\,(\ref{eq:MGcond}) contains 
three relations with $(ij)=(12,23,13)$, 
two of which are independent.

As in the case of four-dimensions, the $D$-dimensional extension of the SM
without supersymmetry (SUSY) does not realize perturbative gauge unification,
unless additional fields are added to the SM particle
spectrum\,\cite{powerlaw}.  
The simplest example of the perturbative gauge
unification is the minimal supersymmetric extension of the SM
(MSSM).  
Following \cite{Ant,powerlaw,Hill-GUT}, we will consider a
$D$-dimensional MSSM compactified on $T^\d=(S^1 /\ZZ_2)^\d$, with
vector-supermultiplets and 
two Higgs supermultiplets propagating in the bulk,
and chiral-supermultiplets for fermions sitting on the brane. For the
simplest GUT group $SU(5)$\,\cite{SU5}, the SM fermions fill an entire
$SU(5)$ representation, their presence in the bulk would not effect
unification, but does change the value of the unified coupling. 
Since matter contributes positively to the beta functions and 
drives the theory to stronger coupling,
allowing matter in the bulk will only enhance the effect of KK
excitations and strengthen our unitarity bounds.  For this $D$-dimensional
MSSM, the coefficients of one-loop beta functions are,
$(b_1,\,b_2,\,b_3)=(33/5,\,1,\,-3)$, and
$(\bt_1,\,\bt_2,\,\bt_3)=(3/5,\,-3,\,-6)$.\, Substituting these into
(\ref{eq:MGcond}), we derive the numerical relation
\beq
\label{eq:MGUT-MSSM}
\ba{ll}
\MG\!\! 
& \dis \simeq\! 10^4\!\exp\!\[28.3
  \!\(1-\f{\,F(\d,\NKK)\,}{44.1}+0.040\D_{12}\)\]{\rm GeV},
\ea
\eeq
where we have imposed the condition
\,$\a_1(\MG)=\a_2(\MG)$\, since the $Z$-pole values of 
$\a_{1,2}$ are much more precisely known than the
value of $\a_3$.
 
%
%
%\vspace*{-5mm}
\begin{figure}[h]
\vspace*{-2mm}
\hspace*{-2.5mm}
\label{fig:Fig1}
%\begin{center}
\includegraphics[width=9.9cm,height=12cm]{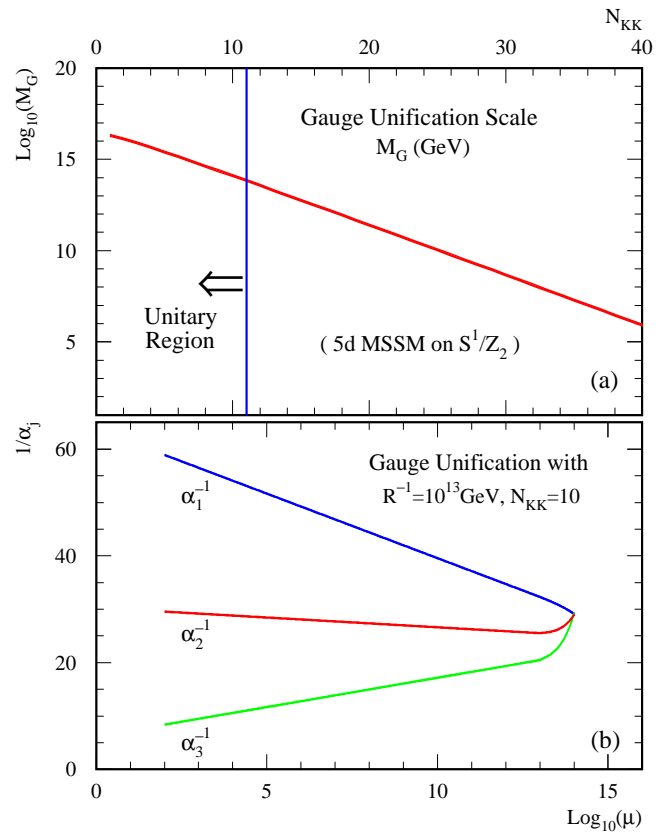} %\hspace*{8mm}
\vspace*{-11mm}
\caption{(a). The GUT scale $\MG$ as a function of
required $\NKK$ value for the 5d MSSM compactified on $S^1/\ZZ_2$. 
Note that, the bound \,$\NKK \leqq 11$\, in 
Eq.\,(\ref{eq:NKK-4dew}) restricts $\MG \gtrsim 10^{14}\,{\rm GeV}$.
(b). The evolution of gauge couplings with the scale $\mu$ (in GeV)
in the 5d MSSM, where $R^{-1}=10^{13}$\,GeV,  $\NKK=10$ and the
unification occurs at $\mu =10^{14}$\,GeV.  
}
%\end{center}
\vspace*{3.5mm}
\end{figure}

Using (\ref{eq:MGUT-MSSM}), we plot the GUT scale $\MG$ as a function of
$\NKK$ for the 5d MSSM in Fig.\,1(a).  In \cite{Hill-GUT}, it was argued that
the term $\D_{ij}$ may receive contributions from the higher dimensional
operators suppressed by the string scale 
which is of ${\cal O}(10^{-2})$ in the
MSSM and could be as large as $10\%$ in its next-to-minimal extension
(NMSSM).  In Fig.\,1(a), we have varied $\D_{12}$ from $+10\%$ to $-10\%$. 
Using the Eq.\,(\ref{eq:gGUT}), 
we also plot in Fig.\,1(b) the evolution of three gauge
coupling constants with a typical high compactification scale
\,$R^{-1}=10^{13}$\,GeV,\, where the higher order parameter $\kappa_{j}$ is
varied within $\pm 10\%$.  
In this case, the unification is accelerated to a
lower scale $\MG=10^{14}$\,GeV
due to the presence of \,$\NKK=10$\, KK states.

For a compactification size $R$ and a GUT scale $\MG$, perturbative
unification can only occur if the KK modes of level $\NKK =R
\MG$ satisfy the unitarity bounds in Sec.\,3.  We note that when $\MG$ is
close to the conventional GUT scale, 
the running coupling \,$\a_2^{~}=g^2/4\pi$\, 
in the range between $R^{-1}$ and $\MG$
has about the same size as (or slightly larger than) the
weak scale value $\alpha_2(m_z)$  [cf. Fig.\,1(b)].  
Hence, we can
apply the unitarity limit \,$\NKK \leqq 11$\, in
Eq.\,(\ref{eq:NKK-4dew}), and find that, because of \,$\NKK =\MG R$,\,
the scales $R^{-1}$ and $\MG$ cannot be separated by more than
one order of magnitude.  This is a generic feature for any
higher-dimensional GUT theory, when the bound (\ref{eq:NKK-4dew}) can
be applied directly.  Therefore, no substantial acceleration of
four-dimensional perturbative gauge unification is possible from
embedding the theory in higher dimensions. In particular, the GUT
scale of the 5d minimal SUSY GUT has to be of order 
$10^{14}$\,GeV or higher
[cf. Fig.\,1(a)].
An extension of our analysis to the 5d GUTs broken by 
orbifolds\,\cite{Hall} can be similarly 
performed\,\cite{5DSM:2001}.

%\vspace*{4.5mm}
\noindent   %\underline
{\bf 5. Conclusions }
\vspace*{1mm}

In this Letter we have extended the study of the unitarity of
compactified 5d Yang-Mills theories, and investigated the unitarity of
the standard model (SM) in higher dimensions.  Analyzing 
gauge-boson scattering in the uncompactified $D$-dimensional theory,
we derive an upper bound on the UV cutoff of the theory.  Using this
estimate in the compactified case yields bounds on the highest
Kaluza-Klein (KK) level $\NKK$ and the Higgs mass $\mH$ allowed in the
effective 4d low-energy theory.  We demonstrated the validity of these
general observations by explicitly analyzing the effective 4d
Kaluza-Klein theory from a compactified five-dimensional SM on
$S^1/\ZZ_2$.  With the analyses before and after compactification, we
derive the strongest bounds on $\NKK$ from the gauge KK scattering.
Applying these bounds to higher-dimensional supersymmetric GUTs, we
show that only a small number of KK states can be used to accelerate
the perturbative gauge coupling unification.  In particular, the
unification scale $\MG$ in the 5d minimal SUSY GUT and similar
theories cannot be lower than about $10^{14}$\,GeV.

As this work was being completed a related study\,\cite{xx} appeared,
which considered bounds on $\mH$ from the scattering of
$W^+_{0L}W^-_{0L}$ into the scalar KK states by using the equivalence
theorem.

%%%%%%%%%%%%%%%%%%%%%%%%%%%%%%%%%%%%%%%%%%%%%%%%%%%%%%%%%%%%%%%%%%%%%%
%\vspace*{2mm}
%\noindent
%{\bf Acknowledgments}\\[2mm]
We thank N. Arkani-Hamed, W. A. Bardeen,
J.~Blum, B. A.~Dobrescu, C. T. Hill, C.~Macesanu, 
and\, W.~W. Repko for discussions. 
This work was supported by the DOE %Department of Energy 
under grants DE-FG02-91ER40676, DE-FG03-93ER40757, 
DE-FG03-98ER41076 and DE-FG02-01ER 45684.

%%%%%%%%%%%%%%%%%%%%%%%%%%%%%%%%%%%%%%%%%%%%%%%%%%%%%%%%%%%%%%%%%%%%%
%\vspace*{-4mm}

\end{document}